\documentclass[amsmath,amssymb,aps,10pt,prd,letterpaper,nofootinbib,balancelastpage,notitlepage,superscriptaddress,twocolumn,floatfix]{revtex4-2}

\usepackage{graphicx}	
\usepackage{epstopdf}	
\usepackage[sort&compress]{natbib}	 	
	\setcitestyle{square,numbers,comma}	
\usepackage[colorlinks=true,urlcolor=blue,linkcolor=black,citecolor=red]{hyperref}
\usepackage{verbatim}

\newcommand{\avg}[1]{\left \langle #1 \right \rangle} 
\newcommand{\abs}[1]{\left | #1 \right |} 
\newcommand{\eq}[1]{\begin{equation} #1 \end{equation}} 
\newcommand{\spliteq}[1]{\begin{equation} \begin{split} #1 \end{split} \end{equation}} 
\newcommand{\parens}[1]{\left ( #1 \right )} 
\newcommand{\brackets}[1]{\left [ #1 \right ]} 
\newcommand{\eqrefTemp}[1]{Eq.~\ref{#1}}
\renewcommand{\eqref}{\eqrefTemp}
\newcommand{\secref}[1]{Sec.~\ref{#1}}
\newcommand{\figref}[1]{Fig.~\ref{#1}}

\newcommand{\unitVec}[1]{\mathbf{\hat{#1}}} 

\newcommand{\Qavg}[1]{\avg{Q^{#1}(\unitVec{n})}}
\newcommand{\Uavg}[1]{\avg{U^{#1}(\unitVec{n})}}

\newcommand{\favg}{\avg{f(t)}}

\newcommand{\Ditau}[1]{\bar{D}_i^{#1}(\unitVec{n},\tau)}

\newcommand{\ritau}[1]{\bar{r}_i^{#1}(\unitVec{n},\tau)}
\newcommand{\rinutau}[2]{\bar{r}_{i,#2}^{#1}(\unitVec{n},\tau)}

\newcommand{\Qcoadd}[1]{\bar{Q}^{#1}(\unitVec{n})}
\newcommand{\Ucoadd}[1]{\bar{U}^{#1}(\unitVec{n})}

\newcommand{\witau}[1]{w_i^{#1}(\unitVec{n},\tau)}
\newcommand{\vitau}[1]{v_i^{#1}(\unitVec{n},\tau)}
\newcommand{\vinutau}[2]{v_{i,#2}^{#1}(\unitVec{n},\tau)}

\newcommand{\Rel}[2]{R^{(#1,#2)}(\tau)}

\newcommand{\Rsrsr}{\Rel{s_{r_1}}{s_{r_2}}}

\newcommand{\fhattype}[1]{\hat{f}^{(#1)}(\tau)}

\newcommand{\fhatdyn}{\fhattype{\mathrm{dyn}}}
\newcommand{\fhatbkg}{\fhattype{\mathrm{bkg}}}

\newcommand{\fjk}{\hat{f}^{(\mathrm{jk})}(\tau)}

\newcommand{\Ajk}{\hat{A}_m^{(\mathrm{jk})}}

\newcommand{\Nitau}{N_i(\unitVec{n},\tau)}
\newcommand{\Sitau}{S_i(\unitVec{n},\tau)}

\begin{document}


\title{BICEP / \emph{Keck} XIV: Improved constraints on axion-like polarization oscillations in the cosmic microwave background}



\author{BICEP/\emph{Keck} Collaboration: P.~A.~R.~Ade}
\affiliation{School of Physics and Astronomy, Cardiff University, Cardiff, CF24 3AA, United Kingdom}
\author{Z.~Ahmed}
\affiliation{Kavli Institute for Particle Astrophysics and Cosmology, SLAC National Accelerator Laboratory, 2575 Sand Hill Rd, Menlo Park, California 94025, USA}
\author{M.~Amiri}
\affiliation{Department of Physics and Astronomy, University of British Columbia, Vancouver, British Columbia, V6T 1Z1, Canada}
\author{D.~Barkats}
\affiliation{Center for Astrophysics, Harvard \& Smithsonian, Cambridge, MA 02138, U.S.A}
\author{R.~Basu Thakur}
\affiliation{Department of Physics, California Institute of Technology, Pasadena, California 91125, USA}
\author{C.~A.~Bischoff}
\affiliation{Department of Physics, University of Cincinnati, Cincinnati, Ohio 45221, USA}
\author{D.~Beck}
\affiliation{Kavli Institute for Particle Astrophysics and Cosmology, SLAC National Accelerator Laboratory, 2575 Sand Hill Rd, Menlo Park, California 94025, USA}
\affiliation{Department of Physics, Stanford University, Stanford, California 94305, USA}
\author{J.~J.~Bock}
\affiliation{Department of Physics, California Institute of Technology, Pasadena, California 91125, USA}
\affiliation{Jet Propulsion Laboratory, Pasadena, California 91109, USA}
\author{H.~Boenish}
\affiliation{Center for Astrophysics, Harvard \& Smithsonian, Cambridge, MA 02138, U.S.A}
\author{E.~Bullock}
\affiliation{Minnesota Institute for Astrophysics, University of Minnesota, Minneapolis, Minnesota 55455, USA}
\author{V.~Buza}
\affiliation{Kavli Institute for Cosmological Physics, University of Chicago, Chicago, IL 60637, USA}
\author{J.~R.~Cheshire IV}
\affiliation{Minnesota Institute for Astrophysics, University of Minnesota, Minneapolis, Minnesota 55455, USA}
\author{J.~Connors}
\affiliation{Center for Astrophysics, Harvard \& Smithsonian, Cambridge, MA 02138, U.S.A}
\author{J.~Cornelison}
\affiliation{Center for Astrophysics, Harvard \& Smithsonian, Cambridge, MA 02138, U.S.A}
\author{M.~Crumrine}
\affiliation{School of Physics and Astronomy, University of Minnesota, Minneapolis, Minnesota 55455, USA}
\author{A.~Cukierman}
\email[Corresponding author: A.~Cukierman\\]{ajcukier@stanford.edu}
\affiliation{Department of Physics, Stanford University, Stanford, California 94305, USA}
\affiliation{Kavli Institute for Particle Astrophysics and Cosmology, SLAC National Accelerator Laboratory, 2575 Sand Hill Rd, Menlo Park, California 94025, USA}
\author{E.~V.~Denison}
\affiliation{National Institute of Standards and Technology, Boulder, Colorado 80305, USA}
\author{M.~Dierickx}
\affiliation{Center for Astrophysics, Harvard \& Smithsonian, Cambridge, MA 02138, U.S.A}
\author{L.~Duband}
\affiliation{Service des Basses Temp\'{e}ratures, Commissariat \`{a} l'Energie Atomique, 38054 Grenoble, France}
\author{M.~Eiben}
\affiliation{Center for Astrophysics, Harvard \& Smithsonian, Cambridge, MA 02138, U.S.A}
\author{S.~Fatigoni}
\affiliation{Department of Physics and Astronomy, University of British Columbia, Vancouver, British Columbia, V6T 1Z1, Canada}
\author{J.~P.~Filippini}
\affiliation{Department of Physics, University of Illinois at Urbana-Champaign, Urbana, Illinois 61801, USA}
\affiliation{Department of Astronomy, University of Illinois at Urbana-Champaign, Urbana, Illinois 61801, USA}
\author{S.~Fliescher}
\affiliation{School of Physics and Astronomy, University of Minnesota, Minneapolis, Minnesota 55455, USA}
\author{N.~Goeckner-Wald}
\affiliation{Department of Physics, Stanford University, Stanford, California 94305, USA}
\author{D.~C.~Goldfinger}
\affiliation{Center for Astrophysics, Harvard \& Smithsonian, Cambridge, MA 02138, U.S.A}
\author{J.~Grayson}
\affiliation{Department of Physics, Stanford University, Stanford, California 94305, USA}
\author{P.~Grimes}
\affiliation{Center for Astrophysics, Harvard \& Smithsonian, Cambridge, MA 02138, U.S.A}
\author{G.~Hall}
\affiliation{School of Physics and Astronomy, University of Minnesota, Minneapolis, Minnesota 55455, USA}
\author{G. Halal}
\affiliation{Department of Physics, Stanford University, Stanford, California 94305, USA}
\author{M.~Halpern}
\affiliation{Department of Physics and Astronomy, University of British Columbia, Vancouver, British Columbia, V6T 1Z1, Canada}
\author{E.~Hand}
\affiliation{Department of Physics, University of Cincinnati, Cincinnati, Ohio 45221, USA}
\author{S.~Harrison}
\affiliation{Center for Astrophysics, Harvard \& Smithsonian, Cambridge, MA 02138, U.S.A}
\author{S. Henderson}
\affiliation{Kavli Institute for Particle Astrophysics and Cosmology, SLAC National Accelerator Laboratory, 2575 Sand Hill Rd, Menlo Park, California 94025, USA}
\author{S.~R.~Hildebrandt}
\affiliation{Department of Physics, California Institute of Technology, Pasadena, California 91125, USA}
\affiliation{Jet Propulsion Laboratory, Pasadena, California 91109, USA}
\author{G.~C.~Hilton}
\affiliation{National Institute of Standards and Technology, Boulder, Colorado 80305, USA}
\author{J.~Hubmayr}
\affiliation{National Institute of Standards and Technology, Boulder, Colorado 80305, USA}
\author{H.~Hui}
\affiliation{Department of Physics, California Institute of Technology, Pasadena, California 91125, USA}
\author{K.~D.~Irwin}
\affiliation{Department of Physics, Stanford University, Stanford, California 94305, USA}
\affiliation{Kavli Institute for Particle Astrophysics and Cosmology, SLAC National Accelerator Laboratory, 2575 Sand Hill Rd, Menlo Park, California 94025, USA}
\affiliation{National Institute of Standards and Technology, Boulder, Colorado 80305, USA}
\author{J.~Kang}
\affiliation{Department of Physics, California Institute of Technology, Pasadena, California 91125, USA}
\author{K.~S.~Karkare}
\affiliation{Center for Astrophysics, Harvard \& Smithsonian, Cambridge, MA 02138, U.S.A}
\affiliation{Kavli Institute for Cosmological Physics, University of Chicago, Chicago, IL 60637, USA}
\author{E.~Karpel}
\affiliation{Department of Physics, Stanford University, Stanford, California 94305, USA}
\author{S.~Kefeli}
\affiliation{Department of Physics, California Institute of Technology, Pasadena, California 91125, USA}
\author{S.~A.~Kernasovskiy}
\affiliation{Department of Physics, Stanford University, Stanford, California 94305, USA}
\author{J.~M.~Kovac}
\affiliation{Center for Astrophysics, Harvard \& Smithsonian, Cambridge, MA 02138, U.S.A}
\affiliation{Department of Physics, Harvard University, Cambridge, MA 02138, USA}
\author{C.~L.~Kuo}
\affiliation{Department of Physics, Stanford University, Stanford, California 94305, USA}
\affiliation{Kavli Institute for Particle Astrophysics and Cosmology, SLAC National Accelerator Laboratory, 2575 Sand Hill Rd, Menlo Park, California 94025, USA}
\author{K.~Lau}
\affiliation{School of Physics and Astronomy, University of Minnesota, Minneapolis, Minnesota 55455, USA}
\author{E.~M.~Leitch}
\affiliation{Kavli Institute for Cosmological Physics, University of Chicago, Chicago, IL 60637, USA}
\author{A.~Lennox}
\affiliation{Department of Physics, University of Illinois at Urbana-Champaign, Urbana, Illinois 61801, USA}
\author{K.~G.~Megerian}
\affiliation{Jet Propulsion Laboratory, Pasadena, California 91109, USA}
\author{L.~Minutolo}
\affiliation{Department of Physics, California Institute of Technology, Pasadena, California 91125, USA}
\author{L.~Moncelsi}
\affiliation{Department of Physics, California Institute of Technology, Pasadena, California 91125, USA}
\author{Y. Nakato}
\affiliation{Department of Physics, Stanford University, Stanford, California 94305, USA}
\author{T.~Namikawa}
\affiliation{Kavli Institute for the Physics and Mathematics of the Universe (WPI), UTIAS, The University of Tokyo, Kashiwa, Chiba 277-8583, Japan}
\author{H.~T.~Nguyen}
\affiliation{Jet Propulsion Laboratory, Pasadena, California 91109, USA}
\author{R.~O'Brient}
\affiliation{Department of Physics, California Institute of Technology, Pasadena, California 91125, USA}
\affiliation{Jet Propulsion Laboratory, Pasadena, California 91109, USA}
\author{R.~W.~Ogburn~IV}
\affiliation{Department of Physics, Stanford University, Stanford, California 94305, USA}
\affiliation{Kavli Institute for Particle Astrophysics and Cosmology, SLAC National Accelerator Laboratory, 2575 Sand Hill Rd, Menlo Park, California 94025, USA}
\author{S.~Palladino}
\affiliation{Department of Physics, University of Cincinnati, Cincinnati, Ohio 45221, USA}
\author{T.~Prouve}
\affiliation{Service des Basses Temp\'{e}ratures, Commissariat \`{a} l'Energie Atomique, 38054 Grenoble, France}
\author{C.~Pryke}
\affiliation{School of Physics and Astronomy, University of Minnesota, Minneapolis, Minnesota 55455, USA}
\affiliation{Minnesota Institute for Astrophysics, University of Minnesota, Minneapolis, Minnesota 55455, USA}
\author{B.~Racine}
\affiliation{Center for Astrophysics, Harvard \& Smithsonian, Cambridge, MA 02138, U.S.A}
\affiliation{Aix-Marseille  Universit\'{e},  CNRS/IN2P3,  CPPM,  Marseille,  France}
\author{C.~D.~Reintsema}
\affiliation{National Institute of Standards and Technology, Boulder, Colorado 80305, USA}
\author{S.~Richter}
\affiliation{Center for Astrophysics, Harvard \& Smithsonian, Cambridge, MA 02138, U.S.A}
\author{A.~Schillaci}
\affiliation{Department of Physics, California Institute of Technology, Pasadena, California 91125, USA}
\author{R.~Schwarz}
\affiliation{School of Physics and Astronomy, University of Minnesota, Minneapolis, Minnesota 55455, USA}
\author{B.~L.~Schmitt}
\affiliation{Center for Astrophysics, Harvard \& Smithsonian, Cambridge, MA 02138, U.S.A}
\author{C.~D.~Sheehy}
\affiliation{Physics Department, Brookhaven National Laboratory, Upton, NY 11973}
\author{A.~Soliman}
\affiliation{Department of Physics, California Institute of Technology, Pasadena, California 91125, USA}
\author{T.~St.~Germaine}
\affiliation{Center for Astrophysics, Harvard \& Smithsonian, Cambridge, MA 02138, U.S.A}
\affiliation{Department of Physics, Harvard University, Cambridge, MA 02138, USA}
\author{B.~Steinbach}
\affiliation{Department of Physics, California Institute of Technology, Pasadena, California 91125, USA}
\author{R.~V.~Sudiwala}
\affiliation{School of Physics and Astronomy, Cardiff University, Cardiff, CF24 3AA, United Kingdom}
\author{G.~P.~Teply}
\affiliation{Department of Physics, California Institute of Technology, Pasadena, California 91125, USA}
\author{K.~L.~Thompson}
\affiliation{Department of Physics, Stanford University, Stanford, California 94305, USA}
\affiliation{Kavli Institute for Particle Astrophysics and Cosmology, SLAC National Accelerator Laboratory, 2575 Sand Hill Rd, Menlo Park, California 94025, USA}
\author{J.~E.~Tolan}
\affiliation{Department of Physics, Stanford University, Stanford, California 94305, USA}
\author{C.~Tucker}
\affiliation{School of Physics and Astronomy, Cardiff University, Cardiff, CF24 3AA, United Kingdom}
\author{A.~D.~Turner}
\affiliation{Jet Propulsion Laboratory, Pasadena, California 91109, USA}
\author{C.~Umilt\`{a}}
\affiliation{Department of Physics, University of Illinois at Urbana-Champaign, Urbana, Illinois 61801, USA}
\author{C.~Verg\`{e}s}
\affiliation{Center for Astrophysics, Harvard \& Smithsonian, Cambridge, MA 02138, U.S.A}
\author{A.~G.~Vieregg}
\affiliation{Department of Physics, Enrico Fermi Institute, University of Chicago, Chicago, IL 60637, USA}
\affiliation{Kavli Institute for Cosmological Physics, University of Chicago, Chicago, IL 60637, USA}
\author{A.~Wandui}
\affiliation{Department of Physics, California Institute of Technology, Pasadena, California 91125, USA}
\author{A.~C.~Weber}
\affiliation{Jet Propulsion Laboratory, Pasadena, California 91109, USA}
\author{D.~V.~Wiebe}
\affiliation{Department of Physics and Astronomy, University of British Columbia, Vancouver, British Columbia, V6T 1Z1, Canada}
\author{J.~Willmert}
\affiliation{School of Physics and Astronomy, University of Minnesota, Minneapolis, Minnesota 55455, USA}
\author{C.~L.~Wong}
\affiliation{Center for Astrophysics, Harvard \& Smithsonian, Cambridge, MA 02138, U.S.A}
\affiliation{Department of Physics, Harvard University, Cambridge, MA 02138, USA}
\author{W.~L.~K.~Wu}
\affiliation{Kavli Institute for Particle Astrophysics and Cosmology, SLAC National Accelerator Laboratory, 2575 Sand Hill Rd, Menlo Park, California 94025, USA}
\author{H.~Yang}
\affiliation{Department of Physics, Stanford University, Stanford, California 94305, USA}
\author{K.~W.~Yoon}
\affiliation{Department of Physics, Stanford University, Stanford, California 94305, USA}
\affiliation{Kavli Institute for Particle Astrophysics and Cosmology, SLAC National Accelerator Laboratory, 2575 Sand Hill Rd, Menlo Park, California 94025, USA}
\author{E.~Young}
\affiliation{Department of Physics, Stanford University, Stanford, California 94305, USA}
\affiliation{Kavli Institute for Particle Astrophysics and Cosmology, SLAC National Accelerator Laboratory, 2575 Sand Hill Rd, Menlo Park, California 94025, USA}
\author{C.~Yu}
\affiliation{Department of Physics, Stanford University, Stanford, California 94305, USA}
\author{L.~Zeng}
\affiliation{Center for Astrophysics, Harvard \& Smithsonian, Cambridge, MA 02138, U.S.A}
\author{C.~Zhang}
\affiliation{Department of Physics, California Institute of Technology, Pasadena, California 91125, USA}
\author{S.~Zhang}
\affiliation{Department of Physics, California Institute of Technology, Pasadena, California 91125, USA}


\date{\today}

\begin{abstract}
We present an improved search for axion-like polarization oscillations in the cosmic microwave background (CMB) with observations from the \emph{Keck Array}. An all-sky, temporally sinusoidal rotation of CMB polarization, equivalent to a time-variable cosmic birefringence, is an observable manifestation of a local axion field and potentially allows a CMB polarimeter to detect axion-like dark matter directly. We describe improvements to the method presented in previous work, and we demonstrate the updated method with an expanded dataset consisting of the 2012-2015 observing seasons. We set limits on the axion-photon coupling constant for mass~$m$ in the range $10^{-23}$-$10^{-18}~\mathrm{eV}$, which corresponds to oscillation periods on the order of hours to years. Our results are consistent with the background model. For periods between~$1$ and~$30~\mathrm{d}$ ($1.6 \times 10^{-21} \leq m \leq 4.8 \times 10^{-20}~\mathrm{eV}$), the $95\%$-confidence upper limits on rotation amplitude are approximately constant with a median of~$0.27^\circ$, which constrains the axion-photon coupling constant to $g_{\phi\gamma} < (4.5 \times 10^{-12}~\mathrm{GeV}^{-1}) m/(10^{-21}~\mathrm{eV}$), if axion-like particles constitute all of the dark matter. More than half of the collected BICEP dataset has yet to be analyzed, and several current and future CMB polarimetry experiments can apply the methods presented here to achieve comparable or superior constraints. In the coming years, oscillation measurements can achieve the sensitivity to rule out unexplored regions of the axion parameter space. 
\end{abstract}


\maketitle

\section{Introduction \label{sec:intro}}

We recently presented constraints on axion-like dark matter from a search for polarization oscillations in the cosmic microwave background (CMB)~\cite{BKXII} (hereafter ``BK-XII''). Here we update the method with several improvements and expand the data volume by a factor of approximately~$4$. This paper is to be viewed as a successor to BK-XII, which we reference heavily. In the presentation of results, we follow the format and, in some cases, even the wording of our previous publication.

The axion is a promising dark-matter candidate~\cite{PQ1977Jun,PQ1977Sep,Weinberg1978,Wilczek1978,Preskill1982,Abbott1982,Dine1982}. Originally introduced in relation to proposed solutions to the strong $CP$~problem in quantum chromodynamics~(QCD), the QCD axion requires a relationship between its mass and its coupling to the Standard Model~(SM). We consider here the much larger class of \emph{axion-like particles}, which are light, bosonic particles with SM~couplings that are similar to those of the QCD axion but without the coupling to QCD. They are, therefore, not related to solutions to the strong $CP$~problem and generally have no fixed relationship between their mass and their SM~coupling. Consequently, they occupy a much larger area in the mass-coupling parameter space. For simplicity, we will hereafter refer to axion-like particles simply as \emph{axions}.

Fedderke et al. proposed two axion observables accessible by current and future CMB polarimetry experiments~\cite{Fedderke2019}. Both are sensitive to the axion-photon coupling constant~$g_{\phi\gamma}$. The first is an overall suppression of CMB polarization, which is referred to as the \emph{washout effect} and which can be constrained by measurements of the $TT$, $TE$ and $EE$~power spectra. The second is a time-varying global rotation of CMB polarization with angular frequency~$m$, the axion mass. The latter observable is called the \emph{AC oscillation} and is the main focus of BK-XII and this work. The AC oscillation can also be considered a form of \emph{time-variable} cosmic birefringence~\cite{Kaufman2014}.

The AC oscillation is a coherent sinusoidal global rotation of CMB polarization with an angular frequency~$m$. Whereas the washout effect is sensitive to axion dark matter present during the epoch of recombination, the AC oscillation is sensitive to axion dark matter at the location of the experiment. The temporal change in CMB polarization is a direct probe of the oscillation of the \emph{local} axion field, and the measurement of axion-like polarization oscillations in the CMB is a form of \emph{direct} dark-matter detection. The oscillation can be taken to be in phase for all CMB experiments and for all photon observing frequencies. For oscillation periods shorter than $\sim 1~\mathrm{d}$, existing axion limits are stronger than what can be achieved with the current generation of CMB instruments (BK-XII Sec.~III~C).
In this work, we consider axion masses in the range of $10^{-23}$-$10^{-18}~\mathrm{eV}$, which roughly corresponds to oscillation periods of hours to years.

Whereas the washout effect, because it relies on the statistics of power spectra, is limited by cosmic variance, the AC oscillation is not. Currently, the washout limits set by Fedderke et al. using publicly available \emph{Planck} power spectra~\cite{Planck2018VI} are within an order of magnitude of the cosmic-variance limit. The oscillation limits presented here (\secref{sec:results}) are less constraining but have not utilized the entirety of the existing BICEP dataset. Furthermore, the constraints from other CMB polarimetry experiments are expected to be comparable to those that can be set with BICEP. Future experiments will achieve even greater sensitivity. In the long term, the oscillation effect, especially when combined across experiments, is likely to become a more sensitive observable than the washout effect. 

Axion-like polarization oscillations can also be observed in astrophysical sources~\cite{Caputo2019,Liu2019,Ivanov2018,Fujita2018,Basu2020}. The axion-photon coupling constant can be constrained through searches for x-ray and gamma-ray anomalies due to astrophysical processes~\cite{Payez2015,Marsh2017,Reynolds2019,Libanov2019,Dessert2020,Calore2020,Carenza2021,Marsh2021} or due to conversion in a strong terrestrial magnetic field~\cite{CAST2017} or through measurements of cosmic distance~\cite{BuenAbad2020}. The mass can be constrained through considerations of small-scale structure~\cite{Irsic2017,Nadler2019,Schutz2020,Nadler2021,Rogers2021}. These probes are complementary and subject to different sets of systematic uncertainties.

In the sections that follow, we rely on the formalism presented in BK-XII. Several general considerations were presented in Sec.~I of BK-XII, and the details of the method were presented in Secs.~III-V and Figs.~1-3. In the current work, we reiterate some but not all of the crucial elements of the method. The current work should be considered a continuation of BK-XII.

The analysis is largely concerned with the time-variable Stokes mixing angle (BK-XII Eq.~3)
\eq{ f(t) \equiv g_{\phi\gamma} \phi_0 \cos(m t + \alpha) , }
where $\phi_0$~is the amplitude of the local axion field (BK-XII Eq.~4) and $\alpha$~is an arbitrary phase. When $g_\mathrm{\phi\gamma} \phi_0 \ll 1$, which is a good approximation in the allowed parameter space, Stokes~$Q$ and~$U$ are rotated into each other by an angle~$f(t)$ (BK-XII Eq.~2) when the on-sky polarization pseudovectors are rotated by an angle~$f(t)/2$. We introduced an estimator~$\hat{f}(\tau)$ (BK-XII Eq.~27) for the Stokes mixing angle, and we use the more concise notation (BK-XII Eq.~41)
\eq{ A \equiv g_{\phi\gamma} \phi_0 \label{eq:A=gphi} }
for the oscillation amplitude. For each value of~$m$ under consideration, we marginalize over~$\alpha$ and set an upper limit on~$A$. Our limits on~$A$ are roughly constant across a broad range of oscillation frequencies. Since $\phi_0 \propto 1/m$, the inferred limits on the axion-photon coupling constant roughly follow $g_{\phi\gamma} \propto m$.

The paper is organized as follows: In \secref{sec:instrument}, we provide a brief overview of the BICEP series of experiments. In \secref{sec:improvements}, we describe the improvements we have made to the method since the publication of BK-XII. In \secref{sec:results}, we present limits on the axion-photon coupling constant from the 2012-2015 observing seasons of the \emph{Keck Array}. This is meant as a first demonstration of the updated method. We close in \secref{sec:conclusions} with expectations and recommendations for current and future CMB polarimetry experiments. 

\section{Instrument overview \label{sec:instrument}}

The \emph{Keck Array} made observations of millimeter-wave polarization from 2012 to~2019 from the South Pole. Five microwave receivers, each similar to the precursor BICEP2~\cite{BK-II}, were integrated on a single mount. The aperture diameter of each receiver is~$25~\mathrm{cm}$. The main observing region occupies $\sim 1\%$ of the sky centered on RA~0h, Dec.~$-57.5^\circ$. In 2012 and 2013, all five receivers operated at~$150~\mathrm{GHz}$. In 2014, two receivers were switched to~$100~\mathrm{GHz}$. In 2015, two $150$-$\mathrm{GHz}$ receivers were switched to~$220~\mathrm{GHz}$. Additional instrument details are provided in~\cite{BKXII,BKSPIDER2015}.

The BICEP series of experiments was designed to search for $B$-mode polarization from primordial gravitational waves~\cite{Kamionkowski1997,Seljak1997} and is also sensitive to $B$~modes created from gravitational lensing and Galactic dust~\cite{Zaldarriaga1998,BKP}. In the most recently released dataset~\cite{BK-X}, the polarization map depths achieved at $100$, $150$ and $220~\mathrm{GHz}$ are, respectively, $5.2$, $2.9$ and~$26~\mu\mathrm{K}_\mathrm{CMB}~\mathrm{arcmin}$. In the results below (\secref{sec:results}), we utilize only a fraction of the dataset and have not exhausted the integrated sensitivity. New results from the BICEP program are forthcoming and will present substantially deeper polarization maps.

Although designed for other purposes, the BICEP instruments, scan strategy and data processing are compatible with an axion-oscillation search. The results below rely on data taken from standard observations targeted at CMB $B$-mode polarization. No change to the scan strategy or low-level data processing is necessary. All frequencies with CMB sensitivity can be used for the axion-oscillation search.

\section{Improvements \label{sec:improvements}}

	In BK-XII~\cite{BKXII}, we described a method of analysis to search for sinusoidal polarization oscillations in BICEP data. We mentioned several potential improvements, which we have now implemented, and we have made a number of additional changes as well. In this section we describe the improvements to the method. Any component of the method that is not discussed here should be assumed to be unchanged from BK-XII.

\subsection{Multiseason dataset \label{sec:multiseason}}
	
	The pipeline is now capable of handling multiple seasons of data from the \emph{Keck Array}. For simplicity and computational speed, the BK-XII analysis focused on a single season, chosen arbitrarily to be 2012. Here we report results for four seasons: 2012-2015. The coadded maps~$\Qcoadd{}$ and~$\Ucoadd{}$ that are used as CMB templates now span multiple seasons and are better representations of the true sky. In Sec.~III~F of BK-XII, we noted that residual noise in the coadded maps dilutes the CMB signal and suppresses the signal transfer function for the Stokes mixing angle~$\hat{f}(\tau)$ (BK-XII Eq.~27). With the single season of BK-XII, this suppression was at the level of~$\sim 30\%$. With the four seasons considered here, the suppression has been reduced to~$\sim 10\%$, and the oscillation signal is more efficiently extracted. As the suppression becomes more negligible, the overall sensitivity to polarization oscillations becomes limited by the noise in each individual scanset rather than residual noise in the coadded maps.
	
	The improvement in the template also sightly increases the background fluctuations~$\fhatbkg$ (BK-XII Eq.~52), but the signal transfer function increases by a greater factor. We consider 
	\eq{ Z(\tau) \equiv \frac{ \fhatdyn }{\sigma(\tau)} \label{eq:Ztau} }
as a measure of the per-scanset signal-to-noise ratio, where $\fhatdyn$~(BK-XII Eq.~50) is a measure of the signal transfer function for scanset~$\tau$ and $\sigma(\tau)$ (BK-XII Eq.~56) is the standard deviation of the background fluctuations. By computing the coadded maps from four seasons instead of one, we find that $Z(\tau)$~increases by approximately~$12\%$.

\subsection{Multifrequency coverage \label{sec:multifrequency}}
	
	The analysis now handles multifrequency data. Beginning in 2014, the \emph{Keck Array} has observed in multiple frequency bands simultaneously (\secref{sec:instrument}). We now produce three sets of coadded maps: one for each frequency band. The oscillation signal is expected in the CMB component of every frequency band.
	
	As discussed in \secref{sec:multiseason}, the template maps~$\ritau{}$ (BK-XII Eq.~20) should be based on the best approximation to the true CMB polarization field. Ideally, a multifrequency component separation could be used to construct the best estimate of the true CMB. For the present analysis, however, this would be superfluous, since the CMB sensitivity is dominated by the $150$-$\mathrm{GHz}$ receivers. See Sec.~I of BK-XII for a discussion of the impact of Galactic foregrounds on axion-oscillation measurements. For foregrounds, we use the Gaussian-dust model described in BK-X~\cite{BK-X}. We opted to use the $150$-$\mathrm{GHz}$ coadded maps as the basis for the rotated-map templates~$\ritau{}$. At~$220~\mathrm{GHz}$, this choice increases the per-scanset signal-to-noise ratio~$Z(\tau)$ by more than a factor of~2, the loss in angular resolution being compensated by a substantial decrease in noise.
	
	We add a frequency subscript to the rotated maps to form~$\rinutau{}{\nu}$, which is to be understood as the rotated map formed from the coadded Stokes parameters for frequency~$\nu$ but with the pointings and polarization orientations for detector~$i$, which is not necessarily active at frequency~$\nu$. We choose $150~\mathrm{GHz}$ to be the template frequency for all detectors. Set $\nu_0 \equiv 150~\mathrm{GHz}$. 
	
	The correlation (BK-XII Eq.~21) becomes
	\eq{ \rho(\tau) \equiv \frac{1}{W(\tau)} \sum_{i,\unitVec{n}} \Ditau{} \rinutau{}{\nu_0} \witau{} \vinutau{}{\nu_0} , }
where $\Ditau{}$~(BK-XII Sec.~III~C) is the pairmap constructed for detector~$i$ during the scanset that occurred at mean time~$\tau$, and $\witau{}$, $\vitau{}$ and $W(\tau)$~are weights defined similarly to Eq.~22 of BK-XII but with $\ritau{} \to \rinutau{}{\nu_0}$ and $\vitau{}$~modified to the pseudo-Wiener form described below in \secref{sec:Wiener}. 

	Similarly, the normalizing constant (BK-XII Eq.~26) becomes
	\eq{ R(\tau) \equiv \frac{1}{W(\tau)} \sum_{i,\unitVec{n}} \rinutau{}{\nu(i)} \rinutau{}{\nu_0} \witau{} \vitau{} , }
where $\nu(i)$~is the frequency at which detector~$i$ observes. For $\nu \not = 150~\mathrm{GHz}$, $R(\tau)$ is a \emph{cross}-correlation rather than an auto-correlation. The cross-correlation is strongest for the CMB component, since the CMB $E$~modes dominate over dust at~$\nu_0 = 150~\mathrm{GHz}$. At~$100~\mathrm{GHz}$, the main non-CMB component is noise, which will not correlate significantly with the $150$-$\mathrm{GHz}$ map. At~$220~\mathrm{GHz}$, in addition to noise, there is substantial dust emission, though it is still subdominant to the CMB $E$~modes as can be seen in, e.g., Fig.~2 of BK-XIII~\cite{BK-XIII}. But neither the noise nor the dust will correlate significantly with the CMB. In Sec.~III~G of BK-XII, we noted a computational speed-up that was made possible by the symmetry of the $\Rsrsr$~matrix (BK-XII Eq.~35). Since the improved analysis allows the template frequency~$\nu_0$ to be different from the detector frequency~$\nu$, the matrix is no longer guaranteed to be symmetric, and all elements must be computed individually.

	The frequency bands have different on-sky beam sizes, and this has \emph{not} been incorporated into the current pipeline. For this reason, the $100$- and $220$-$\mathrm{GHz}$ results are possibly suboptimal. Since the sensitivity of the 2012-2015 analysis is dominated by the $150$-$\mathrm{GHz}$ data, we accept this suboptimality at little cost to the overall constraining power. For future analyses involving nearly equal weights in multiple frequency bands, the beam sizes should be matched between the templates and the pairmaps.

\subsection{Wiener filtering of template maps \label{sec:Wiener}}
	
	In BK-XII, the template weights~$\vitau{}$ were defined (BK-XII Eq.~22) by the inverse noise variance. We noted in Sec.~III~E of BK-XII that, due to the relatively low noise in the $E$-mode coadded maps, this weighting may be suboptimal. We suggested a pseudo-Wiener filter defined by
	\eq{ \vitau{} \equiv \frac{1}{1 + \Nitau/\Sitau} , }
where $\Nitau$~is the noise variance of~$\ritau{}$ and $\Sitau$~is an effective signal level of CMB polarization. In the high-noise limit, we recover an inverse-variance weighting. In the signal-dominated limit, weighting is uniform. We choose a fiducial signal level of $\Sitau = 0.4~\mu\mathrm{K}^2$, which increases~$Z(\tau)$ by approximately~$3\%$. 
 We find that the per-scanset signal-to-noise ratio~$Z(\tau)$ (\eqref{eq:Ztau}) is relatively insensitive to this choice with small-dataset investigations suggesting a roughly~$1\%$ additional increase in~$Z(\tau)$ may be possible.
 
	Together with the improvement from the multiseason coadd (\secref{sec:multiseason}), we find an increase in the per-scanset signal-to-noise ratio~$Z(\tau)$ of roughly~$15\%$. Our final results also benefit from an increase in the total number of scansets. The quantity~$Z(\tau)$ is a measure of the mapping speed rather than the integrated sensitivity. 

\subsection{Incorporation of the oscillation residual \label{sec:oscResidual}}

	We incorporate the effect of the residual oscillation in the coadded maps.
In Sec. III~E~2 of BK-XII, we made the approximation that polarization oscillations average to zero in the coadded maps~$\Qcoadd{}$ and~$\Ucoadd{}$. This allowed us to use a simplified form of the rotated-map template~$\ritau{}$ (BK-XII Eq.~28). The simplified form is convenient, because it does not depend on the oscillation parameters~$A$, $\alpha$ and~$m$, but the approximation is only valid for oscillation periods much shorter than the total observation length. In Sec.~VI~A of BK-XII, we limit the oscillation periods to be no larger than 30~days. 

	We modify Eq.~2 of BK-XII, valid in the limit $f(t) \ll 1$, to be
	\eq{ \parens{ \begin{array}{c} Q(\unitVec{n},t) \\ U(\unitVec{n},t) \end{array} } = \parens{ \begin{array}{cc} 1 & -f(t) \\ f(t) & 1 \end{array} } \parens{ \begin{array}{c} \Qavg{}_t \\ \Uavg{}_t \end{array} } , \label{eq:mixingModified} }
where $\Qavg{}_t$ and $\Uavg{}_t$~denote the \emph{true} time averages, which are to be contrasted with the \emph{observed} averages~$\Qavg{}_\mathrm{obs}$ and~$\Uavg{}_\mathrm{obs}$. The latter are estimated over a finite period of time with interruptions and are to be identified with the measured coadded maps, i.e., observables. The true time averages~$\Qavg{}_t$ and~$\Uavg{}_t$ are only accessible asymptotically. Taking the observation average of \eqref{eq:mixingModified}, we can eliminate~$\Qavg{}_t$ and~$\Uavg{}_t$ to obtain
	\spliteq{ \parens{ \begin{array}{c} Q(\unitVec{n},t) \\ U(\unitVec{n},t) \end{array} } & = \parens{ \begin{array}{cc} 1 & -\brackets{f(t)-\favg_\mathrm{obs}} \\ f(t)-\favg_\mathrm{obs} & 1 \end{array} } \\
& \quad \quad \quad \quad \quad \quad \quad \quad \times \parens{ \begin{array}{c} \Qavg{}_\mathrm{obs} \\ \Uavg{}_\mathrm{obs} \end{array} } , \label{eq:mixingWithObs} }
where $\favg_\mathrm{obs}$~is the average of~$f(t)$ computed with the same weighting that produces~$\Qavg{}_\mathrm{obs}$ and~$\Uavg{}_\mathrm{obs}$. 

	With \eqref{eq:mixingWithObs} in place of Eq.~2 of BK-XII, the analysis can proceed as before with the complication that there is now an \emph{effective} mixing angle
	\eq{ f_\mathrm{eff}(t) \equiv f(t) - \favg_\mathrm{obs} . }
The observation-averaged mixing angle~$\favg_\mathrm{obs}$ must also be computed and depends on the oscillation parameters~$A$, $\alpha$ and~$m$. We estimate~$\favg_\mathrm{obs}$ with
	\eq{ \favg_\mathrm{obs} \approx \frac{1}{W_D} \sum_\tau w_D(\tau) \overline{f}(\tau) , }
where $w_D(\tau)$~is the total weight in pair difference for scanset~$\tau$ and
	\eq{ W_D \equiv \sum_\tau w_D(\tau) . }
Although $\favg_\mathrm{obs}$~depends on oscillation parameters, it need not be completely recomputed for each variation. Instead, for computational efficiency, it can be decomposed into $\tau$-sums that depend only on~$m$. These $m$-dependent $\tau$-sums can be precomputed, so the exploration of the full oscillation parameter space can proceed efficiently.

	The inclusion of the oscillation residual allows the analysis to probe arbitrarily long oscillation periods, though the sensitivity degrades for periods longer than the observation time. For this reason, we limit the analysis to a maximum oscillation period that is similar to the total observation time. At shorter periods, the sensitivity is negligibly altered.

\subsection{Test-statistic distributions}

	We have redefined the jackknife test statistics, so they more closely conform to analytical probability distributions. We do not rely on these probability distributions for estimating $p$~values, instead opting to calibrate through simulations, but it increases confidence in our data model to be able to recover the expectations.

\subsubsection{Nontemporal jackknife tests}

	The nontemporal jackknife tests were introduced in Sec.~V~A of BK-XII. Two mixing angles~$\fhattype{1}$ and~$\fhattype{2}$ are formed from two halves of the data taken during scanset~$\tau$, and the difference~$\fjk$ (BK-XII Eq.~69) is an estimator for systematic contamination. For each~$m$, we formed the quantity~$\Delta q_m^{(\mathrm{jk})}$ (BK-XII Eq.~70), which is a $\Delta\chi^2$~test statistic that compares~$\fjk$ to the \emph{undifferenced} model distribution. We did not compare to the \emph{differenced} model, because the signal transfer function is close to zero. Since the variance of the undifferenced~$\hat{f}(\tau)$ is not equal to the variance of the differenced~$\fjk$, the test statistic~$\Delta q_m^{(\mathrm{jk})}$ was $\chi^2$~distributed only after applying an overall scaling. In the improved analysis, we compare to a \emph{hybrid} model, in which the signal transfer function~$\fhattype{\mathrm{dyn}}$ from the \emph{undifferenced} model is used but the variance is estimated from the \emph{differenced} model. This choice produces a $\Delta q_m^{(\mathrm{jk})}$ test statistic that was verified in simulations to be $\chi^2$~distributed with the expected 2~degrees of freedom.

\subsubsection{Temporal jackknife tests}

	In Sec.~V~B of BK-XII, we introduced the temporal jackknife tests based on the test statistic~$\Ajk$ (BK-XII Eq.~72), which measures the difference in best-fit amplitude~$\hat{A}_m^{(i)}$ between two halves of the data: $i \in \{1,2\}$. We noted that $\Ajk$~is distributed approximately as a one-sided Gaussian. As with the nontemporal jackknife tests, we did not rely on any assumed distribution and instead calibrated through simulations. Since BK-XII, we were able to make the temporal-jackknife test statistic more consistent with Gaussianity by noting that $\hat{A}_m^{(1)} - \hat{A}_m^{(2)}$ does not, in general, vanish and that the mean of background simulations should be subtracted out. Our modified temporal-jackknife statistic is
	\eq{ \Ajk \equiv \abs{ \frac{ \hat{A}_m^{(1)} - \hat{A}_m^{(2)} - \avg{ \hat{A}_m^{(1)} - \hat{A}_m^{(2)} } }{2} } , }
where the mean is to be taken over realizations of the background model. As in BK-XII, we normalize this test statistic by the expected standard deviation (BK-XII Eq.~73) and consider the most extreme value over $m > 0$ (BK-XII Eq.~74).

\subsection{Bias subtraction}

	A small bias has been subtracted from the $\hat{f}(\tau)$~estimator (BK-XII Eq.~27) for the Stokes mixing angle in scanset~$\tau$. In BK-XII, $\hat{f}(\tau)$~was assumed to have a mean of zero. In fact, the non-noise sky components, which are in this case the static CMB and the static dust field, create a bias in~$\hat{f}(\tau)$, which depends on the detector orientations and coverage. This bias is due to the fact that the non-noise components of the rotated map~$\rinutau{}{\nu_0}$ are derived from the same sky as the static components of the per-scanset pairmaps~$\Ditau{}$. The two are approximately but not exactly orthogonal. With the increased sensitivity of four seasons, this bias appears as a spurious oscillation with a period of 4~days, the timescale over which the deck angle of the \emph{Keck Array} is cycled. Other oscillation frequencies are affected negligibly.

	We can estimate the size of the bias by calculating an expectation value based on CMB and dust simulations, and the background model mean is adjusted accordingly.

\subsection{Avoidance frequencies \label{sec:avoidanceFreqs}}

	We have masked the masses~$m$ which correspond to known observation-related timescales. Although we have no evidence for signal-like excess at these oscillation frequencies, we remove them from the analysis preemptively. The removed oscillation periods are $12~\mathrm{h}$, $1~\mathrm{d}$, $2~\mathrm{d}$, $4~\mathrm{d}$ and~$1~\mathrm{yr}$. Neighboring frequencies are correlated, so we remove all frequencies within 3~bins of these. We perform the removal using solar time, but the sidereal values are close enough to be covered by the 3~bins on either side.

\subsection{Season jackknife \label{sec:seasonJackknife}}

	We have introduced a season jackknife which tests for changes in instrumental polarization-angle orientation between seasons. For our standard CMB analysis, we estimate this angle through an $EB$~nulling procedure~\cite{Kaufman2014} with typical values on the order of~$0.2^\circ$. This is at the level of the expected upper limits for the four-season analysis considered here, and we do not perform any $EB$~nulling. If the instrumental polarization rotation changes from season to season, it could mimic an on-sky oscillation with a period of~$1~\mathrm{yr}$ or a related harmonic. We impose a multiseason jackknife test that we call the ``season jackknife". This is a temporal jackknife in the sense of Sec.~V~B of BK-XII. We split the data by alternating seasons, i.e., seasons~1 and~3 together and seasons~2 and~4 together. We perform both the spectrum test and the constant-offset test (BK-XII Sec.~V~B~1). 

\section{Results \label{sec:results}}

With the improvements outlined in \secref{sec:improvements}, we apply our method to the first four observing seasons of the \emph{Keck Array}, which occurred in the years 2012-2015. In BK-XII, we considered only the 2012 season. We chose the first four years of the \emph{Keck Array}, because they represent a modest increase in computational expense over the BK-XII analysis while still testing the multiseason (\secref{sec:multiseason}) and multifrequency (\secref{sec:multifrequency}) features of the improved method.

\subsection{Mass coverage}

	In choosing mass values~$m$ at which to evaluate our upper limits, many of the considerations are similar to those discussed in Sec.~VI~A of BK-XII. We keep the maximum frequency at $\nu_\mathrm{max} = 0.5~\mathrm{hr}^{-1}$ with the associated maximum mass $m_\mathrm{max} = 2 \pi \nu_\mathrm{max}$. The total time range~$T$ that the observations cover is now approximately 4~times \emph{larger}, so the frequency resolution~$\Delta \nu \equiv 1/(\beta T)$ is approximately 4~times \emph{smaller}. We keep the oversampling factor $\beta = 3$. Then we consider the mass range $0 \leq m \leq m_\mathrm{max}$ with resolution $\Delta m = 2 \pi \Delta\nu$. Unlike BK-XII, we impose no low-frequency cutoff due to the incorporation of the oscillation residual (\secref{sec:oscResidual}), but we do remove the observation-synchronized frequencies listed in \secref{sec:avoidanceFreqs}. The lowest non-zero frequency is now $\nu_\mathrm{min} = 2.8 \times 10^{-9}~\mathrm{Hz}$, which corresponds to a minimum mass $m_\mathrm{min} = 1.2 \times 10^{-23}~\mathrm{eV}$.

\subsection{Unblinding procedure \label{sec:unblinding}}

	The unblinding procedure was nearly identical to that described in Sec.~VI~B of BK-XII. The real data products were kept blinded until the jackknife tests were shown to be passed. Since the four seasons of the \emph{Keck Array} considered in this work represent only a fraction of the total BICEP dataset already collected, we have the ability to follow up any signal-like excess with more data. The inclusion of more data, however, is a significant computational expense and not necessary to validate the method improvements outlined in \secref{sec:improvements}. We decided \emph{before unblinding} to set a threshold of~$2.5\sigma$ for the global significance of a signal-like excess which would trigger the analysis of additional data. This corresponds to a global PTE (BK-XII Sec.~IV~D) of $\hat{p} \leq 6.2 \times 10^{-3}$. We measured $\hat{p} = 0.74$ and, therefore, defer the analysis of additional data.

\subsection{Systematics}

	For each of the four seasons, we perform the same systematics tests as those described in Sec.~V of BK-XII. In addition, we perform the season jackknife described in \secref{sec:seasonJackknife}. For each season and also for the multiseason season jackknife, we form the global $p$~values~$\hat{p}^{(\mathrm{sys})}$ and~$\hat{c}^{(\mathrm{sys})}$ (BK-XII Eqs.~75 and~76). There are 10~such $p$~values. We set the requirement that the most extreme value is larger than $0.05/n_\mathrm{tests}$, where $n_\mathrm{tests} = 10$ is the number of tests, and that a Kolmogorov-Smirnov (KS) test for uniformity yields a PTE of $p_\mathrm{KS} \geq 0.05$. The most extreme of the $p$~values is $\hat{p}^{(\mathrm{sys})}(2015) = 0.096$, and we find $p_\mathrm{KS} = 0.37$, so we find no evidence for unmodeled systematic errors in the jackknife tests. 
	
\subsection{Background consistency \label{sec:bkgConsistencyResults}}

\begin{figure}
		\includegraphics[width = 0.5\textwidth]{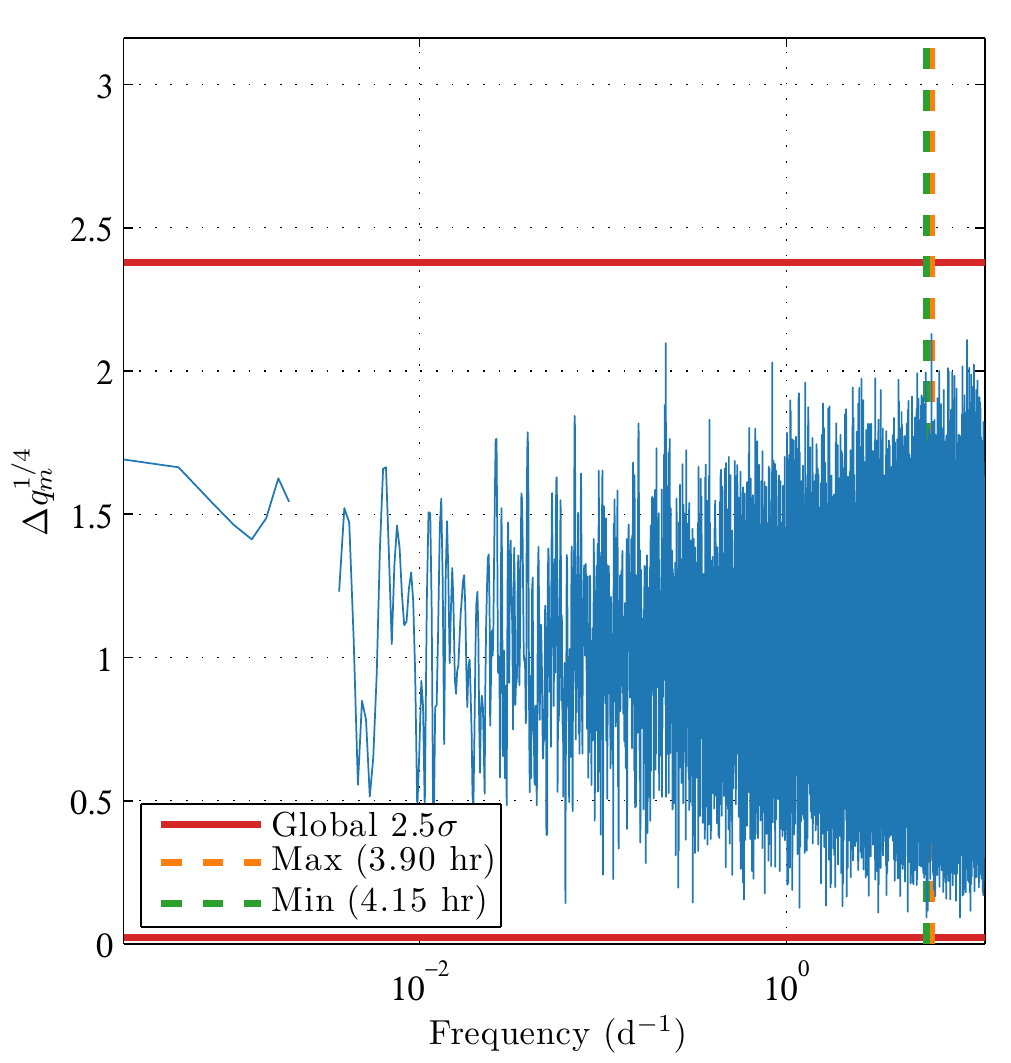}
		\caption{The test statistic~$\Delta q_m$ for consistency with the background model (BK-XII Eq.~67) for real data from the 2012-2015 observing seasons of the \emph{Keck Array}. We plot~$\Delta q_m^{1/4}$ on the vertical axis in order to compress the distribution for visual purposes, and we plot frequency~$m/(2\pi)$ in units of inverse days~($\mathrm{d}^{-1}$) on the horizontal axis. The maximum and minimum values are indicated in the legend with their corresponding oscillation periods. the levels for global $2.5\sigma$~fluctuations in both directions are indicated by horizontal red lines, i.e., there is a $1.2\%$~probability in the background model that at least one value of~$\Delta q_m$ will lie outside the region bounded by the red lines. The visible gap at frequency $1/365~\mathrm{d}^{-1}$ represents an intentional avoidance range (\secref{sec:avoidanceFreqs}). \label{fig:deltaqm}}
	\end{figure}

	As in Sec.~VI~D of BK-XII, we consider the quantity~$\Delta q_m$, which is a $\Delta\chi^2$~test statistic that measures consistency with the background model. Under the null hypothesis, $\Delta q_m$ is $\chi^2$~distributed with $2$~degrees of freedom. In \figref{fig:deltaqm}, we plot~$\Delta q_m$ for real data from the 2012-2015 observing seasons of the \emph{Keck Array}.
There are roughly $5 \times 10^4$~frequency bins in the analysis, so we use~$\Delta\hat{q}$ (BK-XII Eq.~68), the most extreme signal-like value of~$\Delta q_m$, to estimate a global PTE~$\hat{p}$. In \figref{fig:deltaqm}, we plot horizontal lines corresponding to the global $2.5\sigma$ levels for~$\Delta \hat{q}$. All of the data lie within the bounded region. We find $\hat{p} = 0.74$, which lies far below the $2.5\sigma$~requirement to report the results (\secref{sec:unblinding}). Our data are consistent with the background model.

\subsection{Upper limits \label{sec:ULresults}}

\begin{figure*}
		\includegraphics[width = \textwidth]{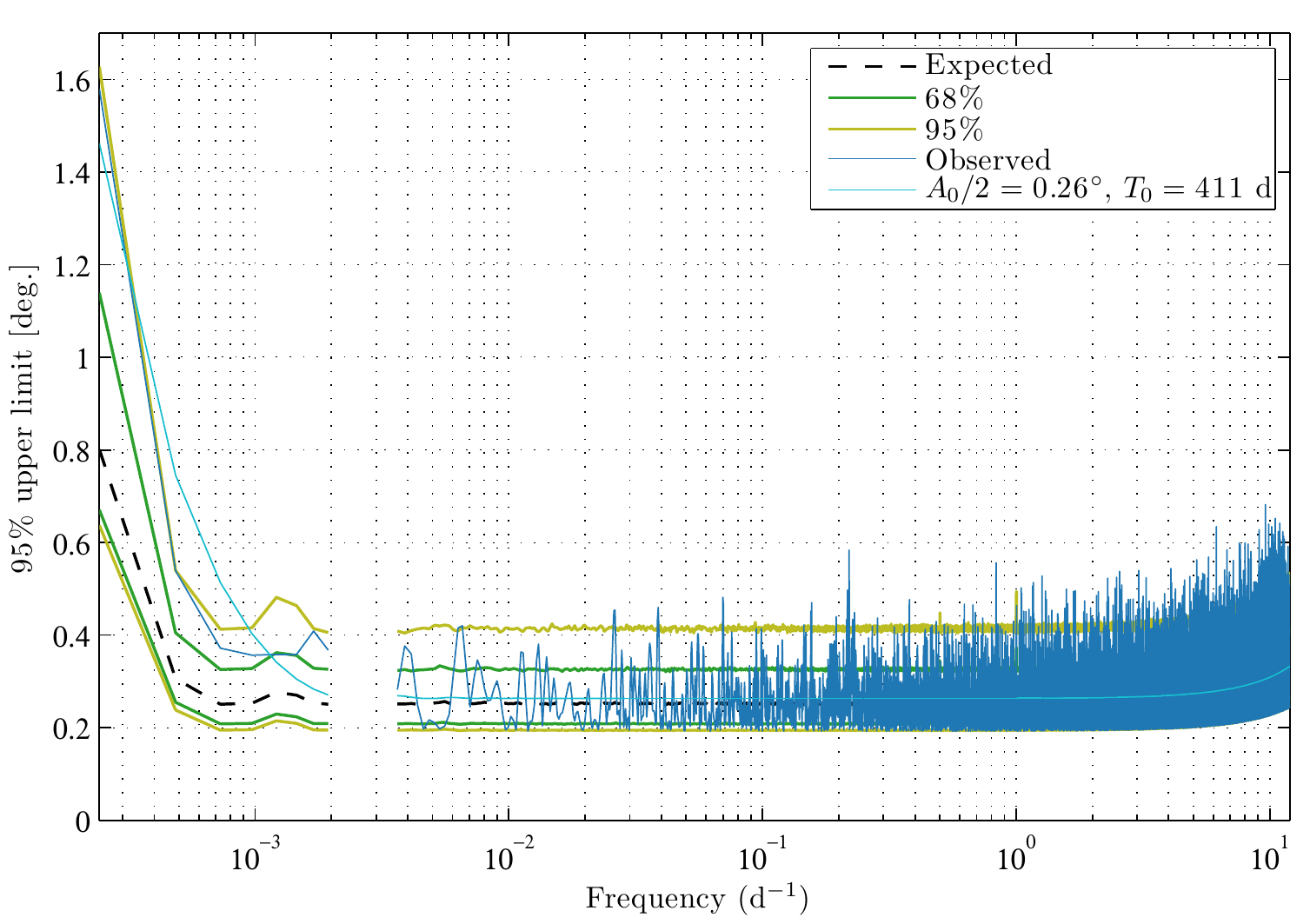}
		\caption{Bayesian $95\%$-confidence upper limits on rotation amplitude~$A/2$ (BK-XII Sec.~IV~C). We also provide the median expectation (black dashes) from background-only simulations as well as $1\sigma$~(green) and $2\sigma$~(yellow) regions. These expectations represent \emph{local} rather than \emph{global} percentiles. With nearly $5 \times 10^4$~frequencies under consideration, we expect several values outside of the $2\sigma$~region. Background consistency is addressed in \secref{sec:bkgConsistencyResults} and \figref{fig:deltaqm}. The median limit for oscillation periods between~$1$ and~$30~\mathrm{d}$ is~$0.27^\circ$. For shorter periods (larger frequencies), the limits are degraded due to binning observations in $\sim 1$-$\mathrm{hr}$ scansets (BK-XII Sec.~III~C). For longer periods (smaller frequencies), the limits are degraded due to the residual oscillation in the coadded maps (\secref{sec:oscResidual}). The breaks in the data are due to the avoidance frequencies discussed in \secref{sec:avoidanceFreqs}. Additionally, we plot a smoothed approximation to our upper limits (\eqref{eq:ulmSmoothed}) in cyan.  \label{fig:ulm}}
	\end{figure*}

	We follow the convention of previous work~(\cite{Fedderke2019,BKXII}) and express our upper limits in terms of the polarization rotation amplitude~$A/2$ rather than the Stokes mixing amplitude~$A$, for which $\hat{f}(\tau)$~is the estimator. We follow Sec.~IV~C of BK-XII to compute $95\%$-confidence upper limits, and we present the results in \figref{fig:ulm}.
For visual comparison, we also overplot the expected distribution of upper limits as implied by background-only simulations. There is a degradation of sensitivity for periods close to the scanset timescale of~$2~\mathrm{hr}$, which is due to the time binning described in Sec.~III~C of BK-XII. For periods longer than~$1~\mathrm{yr}$, we see a more dramatic degradation due to incomplete averaging down of the oscillation residual (\secref{sec:oscResidual}). In BK-XII, we estimated the median limit for periods between~$1$ and~$30~\mathrm{d}$, the region of best and nearly constant sensitivity. The result was $A/2 < 0.68^\circ$ (BK-XII Eq.~77), which is to be compared with the four-season result of
\eq{ A/2 < 0.27^\circ , \label{eq:medianLimit} }
an improvement by a factor of~$2.5$.

	The limits improve by more than the factor of~$\sqrt{4}$ that might be estimated on the basis of the 4-fold increase in the number of seasons. The reason is that several improvements were made to the method (\secref{sec:improvements}) and that the 2012 season, which was used for BK-XII, is actually less sensitive than the 2013 and 2014 seasons. The 2015 season is roughly as sensitive as the 2012 season. 
	
	Over the entire frequency range, we can obtain a smoothed approximation to our upper limits by performing a least-squares fit to (cf. BK-XII Eq.~78)
	\eq{ \frac{A}{2} < \frac{A_0}{2 \operatorname{sinc}\parens{m \Delta t / 2}} \parens{1 - \frac{2}{\parens{T_0 m}^2} \brackets{1 - \cos\parens{T_0 m}}}^{-1/2} \label{eq:ulmSmoothed} } 
with $\Delta t = 44.2~\mathrm{min.}$, which is the median scanset duration, and $A_0$ and $T_0$ as free parameters. We find $A_0/2 = 0.26^\circ$ and $T_0 = 411~\mathrm{d}$. The form of the second factor in \eqref{eq:ulmSmoothed} is due to the oscillation residual (\secref{sec:oscResidual}) and can be derived by assuming continuous, equally weighted observations. It has the effect of degrading the sensitivity for long periods (small~$m$). The $\operatorname{sinc}$ factor is due to the time binning (BK-XII Sec.~III~C) and has the effect of degrading the sensitivity for short periods (large~$m$). We present the smoothed approximation in \eqref{eq:ulmSmoothed} as a convenient form for replotting and comparing with other datasets.

\begin{figure*}
		\includegraphics[width = \textwidth]{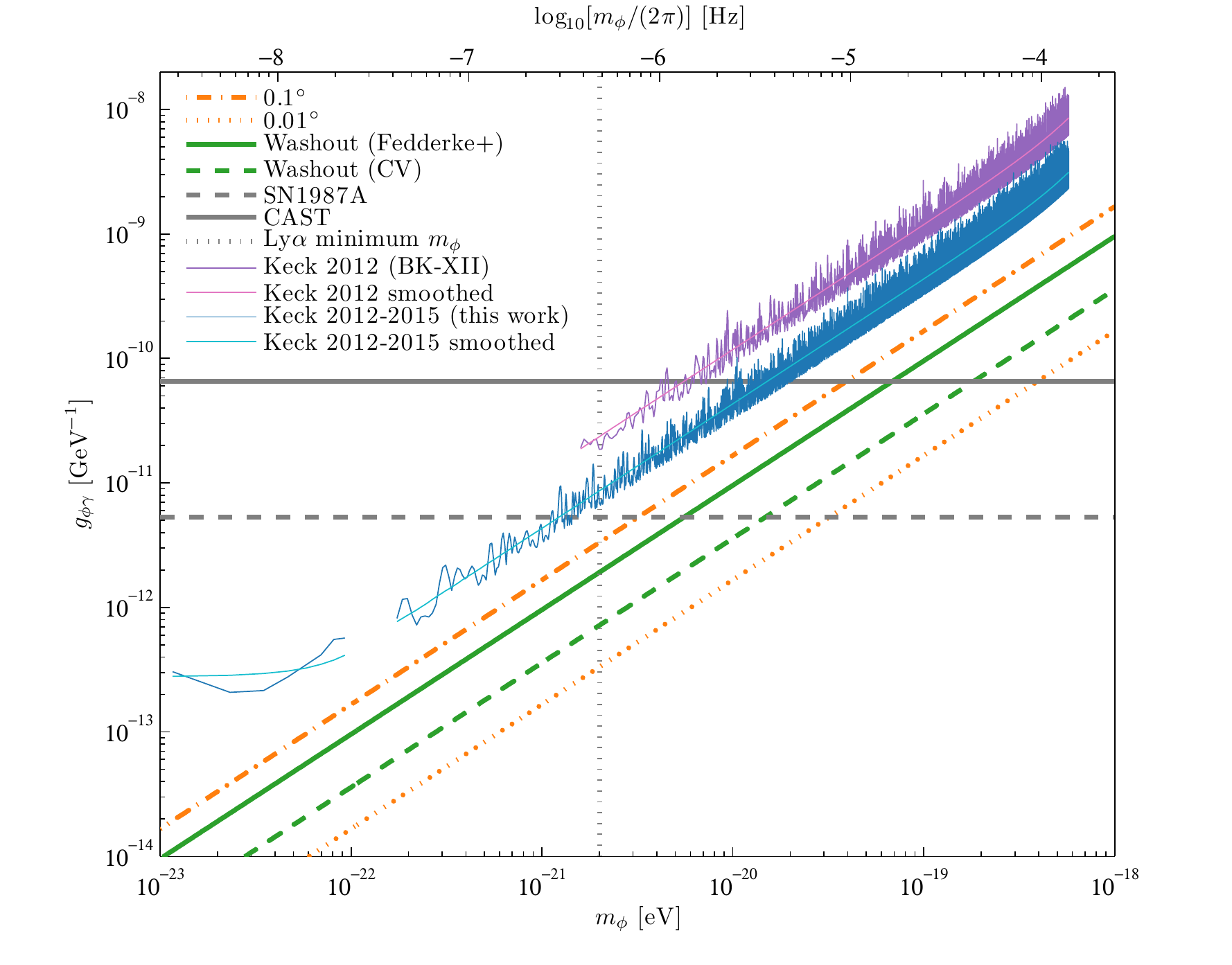}
		\caption{Excluded regions in the mass-coupling parameter space for axion-like dark matter (cf. BK-XII Fig.~6~\cite{BKXII}). All constraints push the allowed regions to larger masses and smaller coupling constants, i.e., toward the bottom right of the figure. If the dark matter is assumed to consist entirely axionlike particles, i.e., if $\kappa = 1$, then our constraints (blue) are immediately implied by Eq.~4 of BK-XII and the results of \figref{fig:ulm}. A smoothed approximation is shown in cyan (\eqref{eq:exclusionSmoothed}). The analogous limits and smoothed approximation from BK-XII are shown in purple and magenta. The orange dot-dashed and dotted lines show the constraints that would be achieved if the rotation amplitude were constrained to~$0.1^\circ$ and~$0.01^\circ$, respectively. The green solid line shows the constraint set by Fedderke et al.~\cite{Fedderke2019} from the washout effect (BK-XII Sec.~I) in \emph{Planck} power spectra. The dashed green line shows the cosmic-variance limit for the washout effect. The dashed grey horizontal line shows the limit from the lack of a gamma-ray excess from SN1987A~\cite{Payez2015}. The solid grey horizontal line is the limit set by the CAST experiment~\cite{CAST2017}. The dotted grey vertical line is a constraint on the minimum axion mass from observations of small-scale structure in the Lyman-$\alpha$ forest~\cite{Irsic2017}, and we note that several similar bounds have also been set by other considerations of small-scale structure~\cite{Nadler2019,Schutz2020,Nadler2021,Rogers2021}.  \label{fig:exclusion}}
	\end{figure*}

	To convert the limits on rotation amplitude to the axion parameter space, we identify $A = g_{\phi\gamma} \phi_0$ (\eqref{eq:A=gphi}).
The limits roughly follow $g_{\phi\gamma} \propto m$ as a result of the $m$~dependence of the axion field strength~$\phi_0$ (BK-XII Eq.~4). In \figref{fig:exclusion}, we present our constraints on the parameter space of axion-like particles from the 2012-2015 observing seasons of the \emph{Keck Array}.
For comparison, we also include the results from BK-XII~\cite{BKXII}, which relied on only the 2012 season. We can transform \eqref{eq:ulmSmoothed} and approximate our upper limits on the coupling constant by (cf. BK-XII Eq.~80)
	\spliteq{ g_{\phi\gamma} & < \parens{ 4.3 \times 10^{-12}~\mathrm{GeV}^{-1} } \operatorname{sinc}^{-1}\parens{\frac{m}{5.0 \times 10^{-19}~\mathrm{eV}}} \\
& \quad \quad \times \parens{\frac{m}{10^{-21}~\mathrm{eV}}} \parens{ \frac{ \kappa \rho_0}{0.3~\mathrm{GeV}/\mathrm{cm}^3} }^{-1/2} \\
& \quad \quad \times \parens{ 1 - 2 \parens{ \frac{m}{m_0}}^{-2} \brackets{ 1 - \cos \parens{ \frac{m}{m_0}}}}^{-1/2} , \label{eq:exclusionSmoothed} }
where $m_0 = 1.9 \times 10^{-23}$,  $\rho_0$~is the local density of dark matter and $\kappa$~is
the fraction of dark matter composed of axion-like particles. As for \eqref{eq:ulmSmoothed}, we provide \eqref{eq:exclusionSmoothed} for convenient replotting and comparison with other constraints.
For periods between~$1$ and~$30~\mathrm{d}$, which corresponds to $1.6 \times 10^{-21} \leq m \leq 4.8 \times 10^{-20}~\mathrm{eV}$, we can transform the limit on the rotation amplitude from \eqref{eq:medianLimit} to obtain
	\spliteq{ g_{\phi\gamma} & < \parens{ 4.5 \times 10^{-12}~\mathrm{GeV}^{-1} } \parens{\frac{m}{10^{-21}~\mathrm{eV}}} \\
& \quad \quad \times \parens{ \frac{ \kappa \rho_0}{0.3~\mathrm{GeV}/\mathrm{cm}^3} }^{-1/2} . }
For comparison, we include in \figref{fig:exclusion} the constraints from other probes. Our constraints from 2012-2015 data do not exclude new regions of parameter space. 

\section{Conclusions and outlook \label{sec:conclusions}}

	We have presented an improved method to search for axion-like polarization oscillations in the CMB, and we have demonstrated the method with data from the 2012-2015 observing seasons of the \emph{Keck Array}. The changes from our previous work~\cite{BKXII} maintain compatibility with the design and operation of experiments targeting primordial $B$~modes, and the search can be continued by current and future projects with no change to scan strategy or to low-level data processing.
	
	The oscillation observable can be considered a form of axion \emph{direct} detection, because it is sensitive to the \emph{local} axion field. In this regard, the oscillation observable is similar to the measurement of CAST, which seeks to directly detect axions by measuring conversion to photons in a strong magnetic field within the experimental apparatus~\cite{CAST2017}. Whereas the CAST constraint in this mass range is nearly constant with~$m$, the oscillation constraints improve as $m$~becomes smaller and are more constraining than CAST for $10^{-23}~\mathrm{eV} \leq m \leq 10^{-20}~\mathrm{eV}$ (\figref{fig:exclusion}).
		
	In addition to the direct measurements of CAST and BICEP/\emph{Keck}, there are a number of indirect constraints from astrophysical and cosmological probes. With only 2012-2015 data from the \emph{Keck Array}, we do not exclude new regions of the axion parameter space, but we note that the polarization-oscillation observable is unique and vulnerable to different systematic biases. Although we have roughly quadrupled the data volume from BK-XII, we have still analyzed less than half of the data collected to date by the BICEP program. Other experiments have been making sensitive measurements as well, and observations will continue with steadily increasing mapping speed.
	
	Already with only four seasons, the residual noise in the coadded polarization maps has become nearly negligible~(\secref{sec:multiseason}). In this limit, the sensitivity to polarization oscillations depends on the number of scansets~$n$, the per-scanset noise and the distribution of scansets in time. For evenly-weighted scansets uniformly distributed in time, the upper limits will scale approximately as~$1/\sqrt{n}$. 
	
	In addition to the four seasons from the \emph{Keck Array} that were analyzed in \secref{sec:results}, the BICEP program has also collected three seasons with BICEP2~\cite{BK-II}, four additional seasons (2016-2019) with the \emph{Keck Array}, five seasons (2016-2020) with BICEP3~\cite{Kang2018} and one season (2020) with the BICEP Array~\cite{Schillaci2020}. The latter will be successively upgraded in coming years to achieve its full design sensitivity.
	
	Sensitivity could improve through time-domain correlations with other CMB experiments. A natural choice for BICEP would be to correlate with the South Pole Telescope (SPT)~\cite{Bender2018}, which is co-located and with which BICEP has already established a formal partnership known as the South Pole Observatory. The BICEP and SPT datasets are complementary in that BICEP has achieved greater integrated polarization sensitivity while SPT has greater angular resolution and, therefore, access to a larger number of polarization modes. 
	
	The CMB Stage-4 (CMB-S4) project~\cite{Carlstrom2019,Abazajian2019} will increase sensitivity to unprecedented levels. The axion-oscillation search described here does not impose requirements on the design or scan strategy of CMB-S4, since the method relies only on repetitive measurements of CMB polarization. Sensitivity is larger at CMB-dominated frequencies like~$95$ and~$150~\mathrm{GHz}$, since the global oscillation affects only the CMB component of the polarization field. All else being equal, higher-resolution observations (aperture diameters of $5$-$10~\mathrm{m}$) are preferable in order to measure the largest number of CMB polarization modes.
	
	The improved method presented in this work can be adapted by other CMB polarimetry experiments. Some of our analysis choices depend on unique characteristics of the \emph{Keck Array} and its dataset, but many of the techniques can be straightforwardly generalized. Simultaneous observations from multiple sites can be combined to protect against systematics and improve overall sensitivity.

\begin{acknowledgments}
The BICEP/\emph{Keck Array} projects have been made possible through a series of grants from the National Science Foundation including 0742818, 0742592, 1044978, 1110087, 1145172, 1145143, 1145248, 1639040, 1638957, 1638978, 1638970 \& 1836010 and by the Keck Foundation.
The development of antenna-coupled detector technology was supported
by the JPL Research and Technology Development Fund and Grants No.\
06-ARPA206-0040 and 10-SAT10-0017 from the NASA APRA and SAT programs.
The development and testing of focal planes were supported
by the Gordon and Betty Moore Foundation at Caltech.
Readout electronics were supported by a Canada Foundation
for Innovation grant to UBC.
The computations in this paper were run on the Odyssey cluster
supported by the FAS Science Division Research Computing Group at
Harvard University.
The analysis effort at Stanford and SLAC was partially supported by the Department of Energy, Contract DE-AC02-76SF00515.
We thank the staff of the U.S. Antarctic Program and in particular
the South Pole Station without whose help this research would not
have been possible.
Most special thanks go to our heroic winter-overs Robert Schwarz
and Steffen Richter.
We thank all those who have contributed past efforts to the BICEP/\emph{Keck Array}
series of experiments.
\end{acknowledgments}

\bibliography{axionOscillations_Keck20122015_PRD}

\end{document}